\documentclass[runningheads]{llncs}
\usepackage[T1]{fontenc}
\usepackage{graphicx}
\newcommand{\orcid}[1]{\href{https://orcid.org/#1}{\includegraphics[width=10pt]{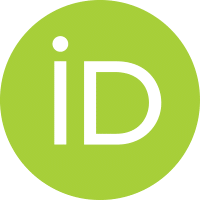}}}
\usepackage{hyperref}
\usepackage{color}

\usepackage[skip=2pt plus1pt]{parskip}
\usepackage{caption}
\usepackage[numbers,sort]{natbib}
\usepackage{multirow}
\usepackage{tabularx}
\usepackage{verbatim}
\usepackage{caption,subcaption}
\usepackage{amsmath}
\usepackage{mdframed}
\usepackage{soul, color, colortbl}

\begin{document}
\title{GRU-Net: Gaussian attention aided dense skip connection based  MultiResUNet for Breast Histopathology Image Segmentation}

\titlerunning{GRU-Net for breast histopathology image segmentation}

\author{Ayush Roy\inst{1}\orcid{0000-0002-9330-6839} \and
Payel Pramanik\inst{2}\orcid{0000-0002-6086-0681} \and
Sohom Ghosal\inst{3}\orcid{0009-0006-3550-1800} \and
Daria Valenkova\inst{4}\orcid{0009-0005-3042-1476} \and
Dmitrii Kaplun\inst{5,4}\orcid{0000-0003-2765-4509} \and
Ram Sarkar\inst{2}\orcid{0000-0001-8813-4086}}

\authorrunning{A. Roy et al.}

\maketitle

\begin{abstract}
Breast cancer is a major global health concern. Pathologists face challenges in analyzing complex features from pathological images, which is a time-consuming and labor-intensive task. Therefore, efficient computer-based diagnostic tools are needed for early detection and treatment planning. This paper presents a modified version of MultiResU-Net for histopathology image segmentation, which is selected as the backbone for its ability to analyze and segment complex features at multiple scales and ensure effective feature flow via skip connections. The modified version also utilizes the Gaussian distribution-based Attention Module (GdAM) to incorporate histopathology-relevant text information in a Gaussian distribution. The sampled features from the Gaussian text feature-guided distribution highlight specific spatial regions based on prior knowledge. Finally, using the Controlled Dense Residual Block (CDRB) on skip connections of MultiResU-Net, the information is transferred from the encoder layers to the decoder layers in a controlled manner using a scaling parameter derived from the extracted spatial features. We validate our approach on two diverse breast cancer histopathology image datasets: TNBC and MonuSeg, demonstrating superior segmentation performance compared to state-of-the-art methods. The code for our proposed model is available on \href{https://github.com/AyushRoy2001/GRU-Net}{GitHub}.
\keywords{MultiResUnet  \and Gaussian Attention \and Deep Learning \and Histopathology \and Text descriptor.}
\end{abstract}

\section{Introduction}
Pathological images play a crucial role in clinical diagnosis, especially for cancer grading. Medical professionals analyze the appearance and morphology of cells, from individual cells to entire tissue slices, to provide qualitative and quantitative assessments. Cell identification and quantitative analysis are vital in pathological examinations, helping medical practitioners identify specific cancer subtypes, evaluate cancer progression stages, and establish connections with genetic mutations. Cell analysis includes various essential tasks such as cell type classification, characterization of cell shapes, and determination of nucleus and cell percentages. In the past, pathologists and medical experts have carried out these tasks manually, including cell segmentation, detection, and classification. However, this traditional approach is extremely labor-intensive and time-consuming. The shortage of pathologists in many regions, especially developing countries, has highlighted the need for alternative and effective solutions. According to Metter et al \cite{metter2019trends}, the current number of pathologists is often insufficient to meet the growing demand for accurate and timely pathological assessments. Hence, innovative technologies are urgently needed to expedite the analysis of pathological images and ease the burdensome tasks that medical professionals currently face.\\

In recent years, there has been a significant exploration of machine learning and deep learning techniques for computer-aided diagnosis of pathological images, offering advantages over conventional segmentation algorithms such as watershed\cite{ranefall1997new}, color-based thresholding\cite{xu2014efficient}, super-pixels\cite{sornapudi2018deep}, level sets\cite{husham2016automated}, and graph cut\cite{qi2014dense}. Deep learning methods, particularly convolutional neural networks (CNNs), including models like fully convolutional network (FCN)\cite{long2015fully}, U-Net\cite{su2022improved}, PSPNet\cite{zhao2017pyramid}, and DeepLab series\cite{chen2014semantic,chen2017rethinking,chen2018encoder}, have shown superior performance in segmenting natural images, which form a strong foundation for their application in medical image segmentation. For instance, Histoseg by Wazir et al.~\cite{wazir2022histoseg} incorporates attention mechanisms for cell segmentation, utilizing attention units for global and local feature representations along with a multi-loss function. Singha et al. ~\cite{singha2023alexsegnet} presented AlexSegNet, a deep learning model inspired by the AlexNet architecture, which employs an encoder-decoder framework and fusion of feature maps in the channel dimension within the encoder, along with skip connections in the decoder to combine low and high-level features, ensuring effective nucleus segmentation. Keaton et al.~\cite{keaton2023celltranspose} introduced CellTranspose, a few-shot domain adaptation approach for cellular instance segmentation across 2D and 3D microscopy datasets. Xia et al.~\cite{xia2023weakly} introduced a weakly supervised nuclei segmentation method that relies solely on point annotations, leveraging dual input information from weakly supervised segmentation and auxiliary colorization tasks to enhance segmentation accuracy. Kanadath et. al~\cite{kanadath2023multilevel} introduced the MMPSO (multilevel multiobjective particle swarm optimization guided superpixel) algorithm, combining multilevel multiobjective particle swarm optimization with superpixel clustering, to identify optimal threshold values in histopathology image segmentation. Whereas Naylor et.~\cite{naylor2018segmentation} suggested an automatic segmentation method of cell nuclei from H\&E stained histopathology data using fully convolutional networks, with a focus on addressing the challenge of segmenting touching nuclei by formulating the problem as a regression task of the distance map. 

\begin{figure}
     \centering
     \subfloat[Histopathology image\label{1a}]{%
      \includegraphics[width=0.45\textwidth]{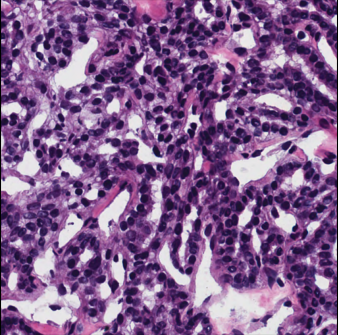}}
     \hfill
      \subfloat[Pixel-level annotations\label{1b}]{%
      \includegraphics[width=0.45\textwidth]{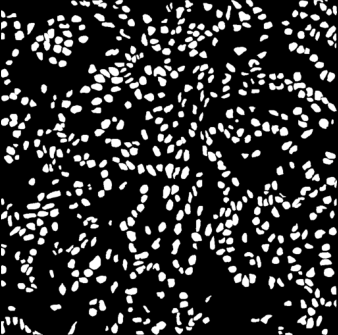}}
      \caption{Example of nuclei segmentation. Sample histopathology image and its pixel-level annotations.}
     \label{fig:intro}
\end{figure}

\subsection{Motivation \& Contributions}
Despite advancements and the availability of resources, challenges persist in the application of deep learning models to clinical cell segmentation. Cell segmentation remains complex due to the intricate and noisy backgrounds present in pathological images, particularly at the cellular level. In this study, we propose a method for accurately segmenting nuclei in histopathology images by leveraging text supervision to decode conditioned semantic information. Our approach also controls the transfer of information from the encoder to the decoder side through skip connections, thereby reducing the transfer of irrelevant features. The main \textbf{contributions} of this work include: 

(i) The backbone used in this work is MultiResU-Net, which has been selected due to its ability to effectively analyze and segment complex features at multiple scales and ensure an effective feature flow via skip connections. Its capability to extract features at various levels makes it an ideal starting point for tasks such as histopathology image segmentation.

(ii) The Gaussian distribution-based Attention Module (GdAM) enables a multi-modal attention mechanism, which incorporates histopathology-relevant text information. This mechanism utilizes the statistical information of the text-encoded features and the spatial features of the bottleneck layer to highlight specific spatial regions based on prior knowledge.

(iii) Using the Controlled Dense Residual Block (CDRB) on skip connections of MultiResU-Net (used as the baseline in our work), the information transfers from the encoder layers ($MRB_E$) to the decoder layers ($MRB_D$) is controlled using a scaling parameter derived from the spatial features extracted by $MRB_E$. 

\begin{figure}[!ht]
\centering
\includegraphics[width=0.8\linewidth]{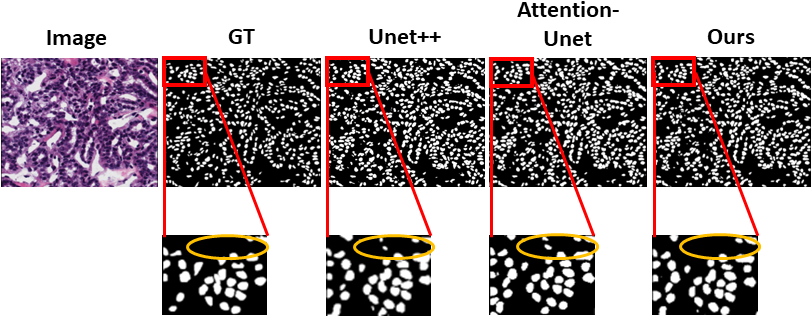}
\caption{Qualitative comparison of the proposed model with Attention-UNet and U-Net++.}
\label{fig:motivation}
\end{figure}

Fig. \ref{fig:motivation} shows the advantage of leveraging textual features and controlling the information transfer from the encoder to the decoder side. It can be seen that Attention-UNet and U-Net++ often over-segment the input image as highlighted by the yellow oval in the zoomed region, whereas the proposed model limits it by leveraging prior knowledge of text labels.

\section{Methodology}
\begin{figure*}[!ht]
\centering
\includegraphics[width=\linewidth]{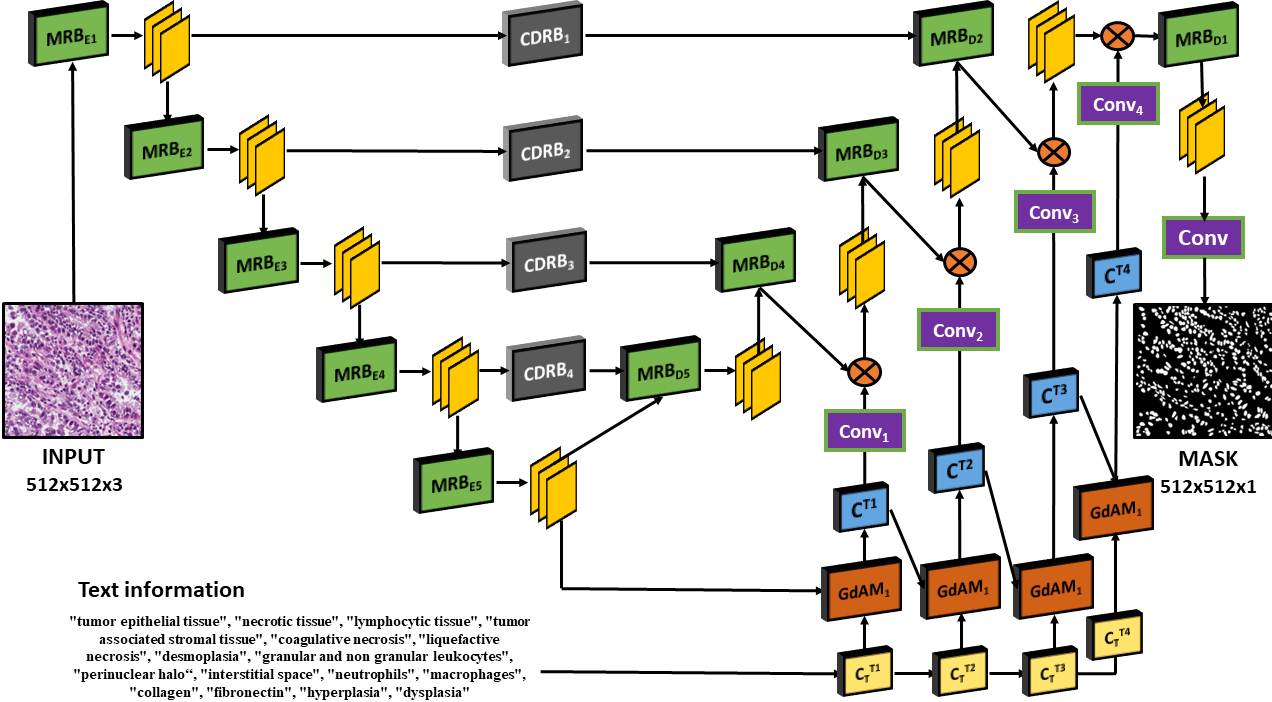}
\caption{Overall workflow of the proposed GRU-Net.}
\label{fig:architecture}
\end{figure*}

Our model, GRU-Net, enhances flexibility in handling histopathology image datasets of varying complexity. Using the Controlled Dense Residual Block (CDRB), the information transfers from the encoder layers ($MRB_E$) to the decoder layers ($MRB_D$) is controlled. A multi-modal attention mechanism incorporating histopathology-relevant text information is achieved using the Gaussian distribution-based Attention Module (GdAM), which utilizes the prior statistical information of the text-encoded features and the spatial features of the bottleneck layer to highlight specific spatial regions conditioned on the prior knowledge. The text prompts used in this work are histopathology-relevant texts and the learned features help enrich the overall distribution from which the attention weights are sampled. A block diagram representation of the proposed model is shown in Fig \ref{fig:architecture}.

\subsection{MultiResUNet}
The MultiResUNet \cite{ibtehaz2020multiresunet} represents an innovative framework tailored for image segmentation and medical image analysis. Its architectural strength lies in the fusion of Multi-Residual Blocks (MRBs) for feature extraction and Residual Blocks (RBs) employed in the skip connections, further advancing upon the foundational principles of the U-Net architecture. 

The effectiveness of MultiResUNet is largely due to the MRBs, which can capture information at multiple scales by combining features from various pathways. As shown in Fig. \ref{fig:intro}, the size and proximity of foreground regions can vary greatly, making it important to integrate multi-scale features that can encompass both local and global contexts. By interacting at different resolutions, this model can capture intricate details and contextual nuances necessary for precise segmentation.

In addition, the presence of residual blocks (RBs) is crucial in enabling the smooth flow of information and gradients between different layers of a deep neural network. This helps to overcome challenges related to gradients during training. By utilizing multiple pathways, the MultiResUNet model can effectively analyze and segment complex medical images, which makes it an excellent starting point for tasks like histopathological image segmentation.

\subsection{Controlled Dense Residual Block}
\begin{figure}[!ht]
\centering
\includegraphics[width=0.8\linewidth]{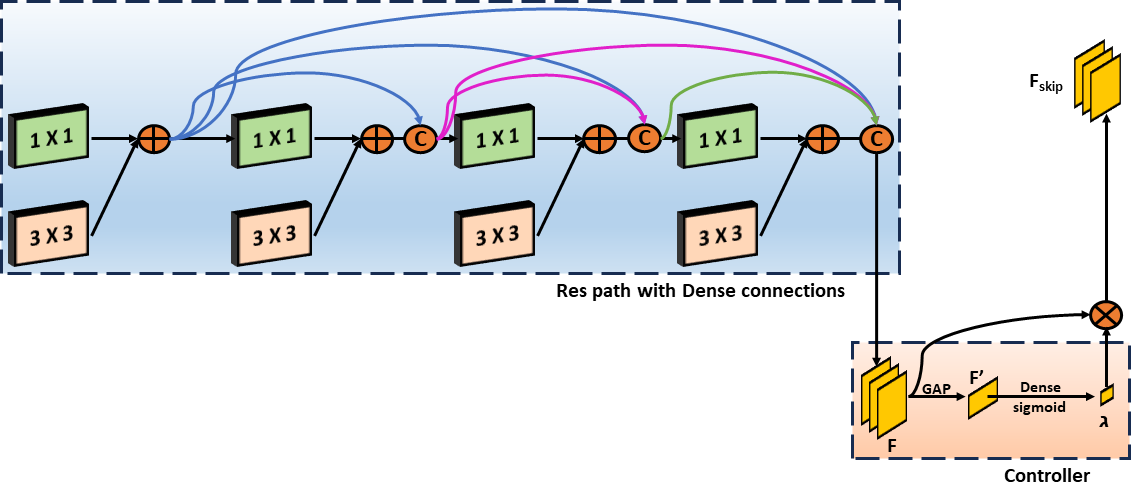}
\caption{Block diagram representation of CDRB module.}
\label{fig:CDRB}
\end{figure}

The Controlled Dense Residual Block (CDRB) is an essential part used in the skip connections between the encoder and the decoder layers, which are known as $MRB_E$ and $MRB_D$, respectively. The CDRB enhances and regulates the information flow between these layers. It consists of two main components: the Res path and the Controller. The block diagram of the CDRB module is shown in Fig.~\ref{fig:CDRB}.

The Res path is a crucial component of the MultiResU-Net that captures multi-scale information needed to achieve accurate segmentation. Res blocks are incorporated into the network architecture, inspired by ResNet designs, making it possible to extract features from different resolutions or scales effectively. This is particularly useful in image segmentation tasks where objects or regions of interest vary significantly in size or context within the image. To ensure that features extracted by earlier layers are preserved and propagated through the network, dense connections are introduced among the components of the Res path. This enhances gradient flow, allowing the network to capture a diverse range of features at different levels of abstraction. As a result, the network can create more comprehensive and discriminative representations.

The controller unit has the important task of controlling the amount of information sent to the decoder layers. This is done to suppress irrelevant information that may negatively impact the decoder's performance. To achieve this, the controller uses a scaling parameter called $\lambda$. The controller takes the output feature from the Res path, denoted as $F$, with dimensions of $B \times H \times W \times C$, and flattens it using Global Average Pooling (GAP) to produce $F'$ with the dimensions of $B \times C$. $F'$ then goes through Dense layers with sigmoid activation to generate $\lambda$, which has dimensions of $B \times 1$. The sigmoid activation function ensures that the value of $\lambda$ ranges from 0 to 1, enabling the scaling of weights to transfer either the entire information (if $\lambda$ = 1), no information (if $\lambda$ = 0), or a fraction of the information (if $\lambda$ is between 0 and 1). Finally, the controller multiplies the scaling factor $\lambda$ with $F$ to produce $F_{skip}$, which contains only the relevant information that the decoder can use.

\subsection{Gaussian distribution-based Attention Module}
Segmentation of histopathology images is a critical task in medical image analysis as it enables identifying and outlining regions of interest, such as tumor tissues, necrotic regions, and inflammatory cells. Traditional segmentation methods rely solely on visual features extracted from images, which may not take into account the contextual information available in textual descriptions associated with histopathological samples. To address this challenge, the GdAM utilizes histopathology-related textual information to enhance the performance of the attention mechanism in histopathology image segmentation. By incorporating statistical features derived from text labels, the model improves the segmentation process by conditioning attention on prior knowledge.

We have a set of carefully selected text labels that represent histopathological features, such as tumor types, tissue characteristics, and cellular components. The text labels include "tumor epithelial tissue," "necrotic tissue," "lymphocytic tissue," "tumor-associated stromal tissue," "coagulative necrosis," "liquefactive necrosis," "desmoplasia," "granular and non-granular leukocytes," "perinuclear halo," "interstitial space," "neutrophils," "macrophages," "collagen," "fibronectin," "hyperplasia," and "dysplasia." We encode these text labels using Distil-BERT~\cite{sanh2019distilbert} into numerical representations, producing a $16 \times N$ matrix, where N is the encoded vector dimension. The matrix is then processed further by Dense layers to convert it to a dimension of $32 \times 32$. We then use GdAM to incorporate the textual embeddings and provide text-guided prior knowledge about relevant histopathological features. 

GdAM incorporates a Gaussian distribution function $\eta (\mu, \sigma)$ that calculates the mean $\mu$ and standard deviation $\sigma$ based on both the input spatial feature map $F$ and the textual information $T$ as shown in Eq.~\ref{eq:1} and ~\ref{eq:2}, where $\mu_T$ and $\mu_F$ are the mean of T and F, respectively and $\sigma_T$ and $\sigma_F$ are the standard deviation of T and F, respectively. 

\begin{equation}
    \mu = \mu_{F} + \mu_{T}
    \label{eq:1}
\end{equation}

\begin{equation}
    \sigma = \sqrt{((\sigma_T)^2 + (\sigma_F)^2)}
    \label{eq:2}
\end{equation}

\begin{figure}[!ht]
\centering
\includegraphics[width=0.8\linewidth]{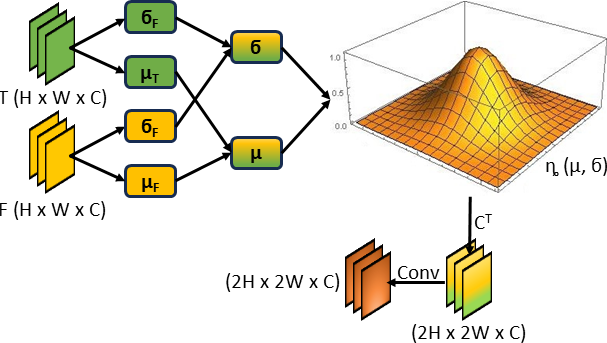}
\caption{Block diagram representation of GdAM module.}
\label{fig:GdAM}
\end{figure}

$\eta (\mu, \sigma)$ is a distribution encoded with the spatial and textual statistical information. This prior knowledge-based distribution is used for extracting information for creating the attention weights. The feature maps extracted from $\eta (\mu, \sigma)$ using Convolution Transpose layers ($C^T$) with rectified linear unit (ReLU) activation function are converted into attention maps using a convolution layer with sigmoid activation (Conv). This process ensures that the segmentation process is guided by both visual features and prior knowledge encoded in the textual information. Fig. \ref{fig:GdAM} demonstrates the overall steps involved in GdAM.

\subsection{Loss Functions}
The Dice loss~\cite{soomro2018strided} and Binary Cross Entropy (BCE) loss~\cite{jadon2020survey} are crucial for image segmentation tasks, evaluating model performance by comparing predicted and actual masks.

The Dice loss, originating from the Dice Coefficient, measures the resemblance between predicted and true masks. It is computed using Eq.~\ref{eq:dice_loss}, where TP (True positive) denotes correctly identified object pixels, TN (True negative) represents correctly identified non-object pixels, FP (False positive) denotes falsely identified object pixels and FN (False negative) represents falsely detected non-object pixels.

\begin{equation}
    \textit{Dice Loss} = 1 - \frac{{2 \times TP}}{{2 \times TP + FP + FN}}
    \label{eq:dice_loss}
\end{equation}

The BCE loss is also widely adopted, quantifying the dissimilarity between predicted and actual masks. Eq.~\ref{eq:bce_loss} formulates it, where $N$ is the pixel count, $y_i$ denotes the true label for pixel $i$, and $p_i$ indicates the predicted foreground class probability.

\begin{equation}
    \textit{BCE Loss} = -\frac{1}{N}\sum_{i=1}^{N}\left(y_i \log(p_i) + (1 - y_i) \log(1 - p_i)\right)
    \label{eq:bce_loss}
\end{equation}

A combination of Dice loss and BCE loss is employed to train our proposed model, which is defined in Eq.~\ref{eq:seg_loss}.

\begin{equation}
    \textit{Hybrid loss} = \textit{BCE Loss} + \textit{Dice Loss}
    \label{eq:seg_loss}
\end{equation}

The performance of our model is evaluated using standard metrics including dice coefficient, Intersection over Union (IoU), precision, and recall, providing quantitative insights into the model's segmentation accuracy.

\section{Results}
\subsection{Experimental Setup}
We have conducted evaluations of our model on two distinct datasets: MonuSeg \cite{kumar2019multi} and TNBC \cite{naylor2018segmentation}. The MonuSeg dataset comprises Hematoxylin and Eosin-stained tissue images, each with dimensions of $512\times512$. It consists of 30 training images, totaling 22,000 annotations, and 14 test images with 7,000 annotations. On the other hand, the TNBC dataset focuses specifically on Triple-Negative Breast Cancer tissues, containing 50 images with 4,022 annotated cells. 

For both datasets, we have utilized the original image size of $512\times512$ as input. Training has been conducted using a split of 70\% for training, 20\% for validation, and 10\% for testing. We have employed a learning rate of 0.0001, utilized the Adam optimizer, set a batch size of 2, and trained the model for 100 epochs. The implementation has been carried out using TensorFlow on an NVIDIA TESLA P100 GPU. This experimental setup has been consistent throughout our ablation study. 

% \subsection{Evaluation metrics}

% \textbf{Accuracy} evaluates overall correctness, calculated as the ratio of correctly classified to total pixels, defined as Eq.~\ref{eq:accuracy}.
% \begin{equation}
%     \text{Accuracy} = \frac{{\text{TP + TN}}}{{\text{TP + TN + FP + FN}}}
%     \label{eq:accuracy}
% \end{equation}

% \textbf{Precision} measures true positive fraction among positives, defined as Eq.~\ref{eq:precision}.
% \begin{equation}
%     \text{Precision} = \frac{{\text{TP}}}{{\text{TP + FP}}}
%     \label{eq:precision}
% \end{equation}

% \textbf{Recall} quantifies true positive rate, defined as Eq.~\ref{eq:recall}.
% \begin{equation}
%     \text{Recall} = \frac{{\text{TP}}}{{\text{TP + FN}}}
%     \label{eq:recall}
% \end{equation}

% \textbf{Intersection over Union (IoU)} gauges overlap between ground truth and predicted masks, defined as Eq.~\ref{eq:iou}.
% \begin{equation}
%     \text{IoU} = \frac{{\text{TP}}}{{\text{TP + FP + FN}}}
%     \label{eq:iou}
% \end{equation}

% \textbf{Dice score}, combining precision and recall, is defined as Eq.~\ref{eq:dice}.
% \begin{equation}
%     \text{Dice score} = \frac{{2 \times \text{TP}}}{{2 \times \text{TP} + \text{FP} + \text{FN}}}
%     \label{eq:dice}
% \end{equation}

\subsection{Ablation study}
To figure out the optimal setup and parameters for our model, we have performed an extensive ablation study on the MonuSeg dataset. The experiments are listed below:

(i) MultiResUnet

(ii) (i) + CDRB without controller

(iii) (i) + CDRB with controller

(iv) The proposed model: (iii) + GdAM

\begin{figure*}[!ht]
     \centering
     \subfloat[Heatmaps for GdAM \label{fig:seg_GdAM_Monuseg}]{%
      \includegraphics[width=0.49\textwidth, height=3.5cm]{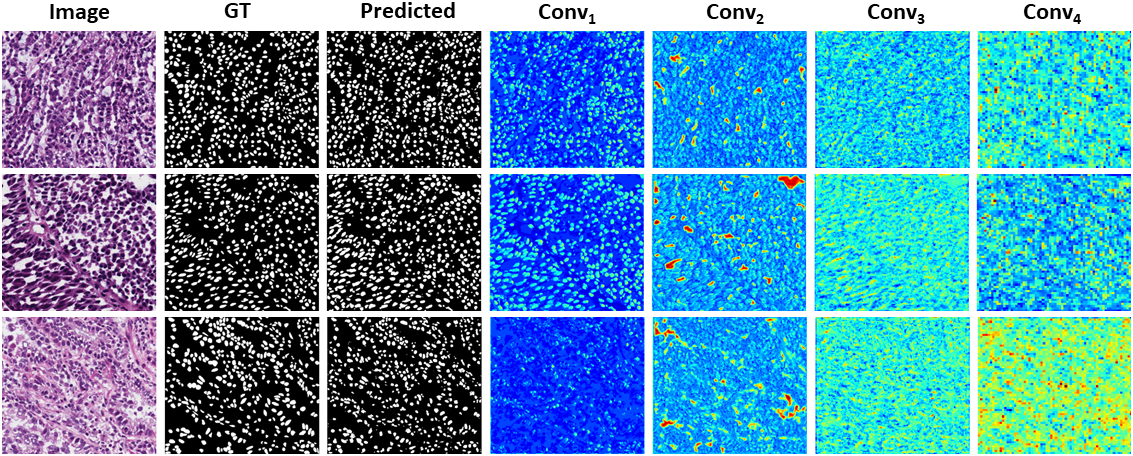}}
     \hfill
      \subfloat[Heatmaps for Encoder and Decoder layers \label{fig:seg_heat_Monuseg}]{%
      \includegraphics[width=0.49\textwidth, height=3.5cm]{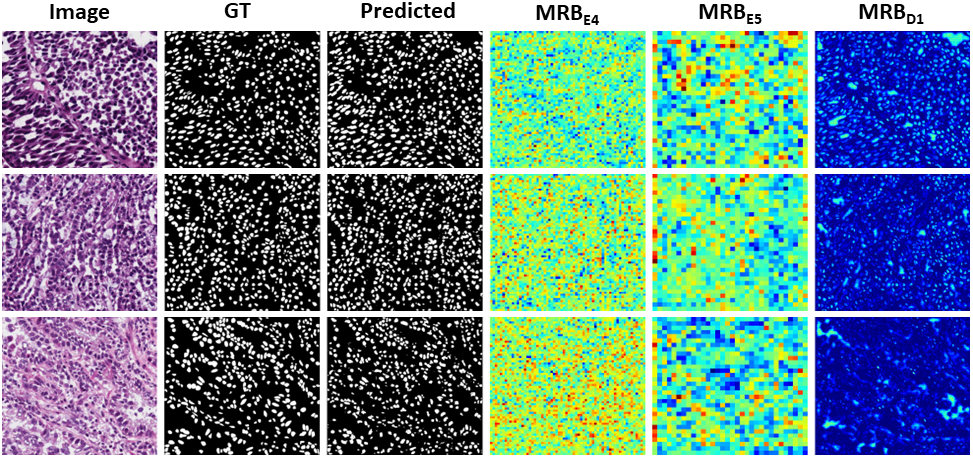}}
      \caption{Segmentation results on the MonuSeg dataset. GT represents ground truth.}
     \label{fig:MonuSeg}
\end{figure*}

\begin{figure*}[!ht]
     \centering
     \subfloat[Heatmaps for GdAM \label{fig:seg_GdAM_TNBC}]{%
      \includegraphics[width=0.49\textwidth, height=3.5cm]{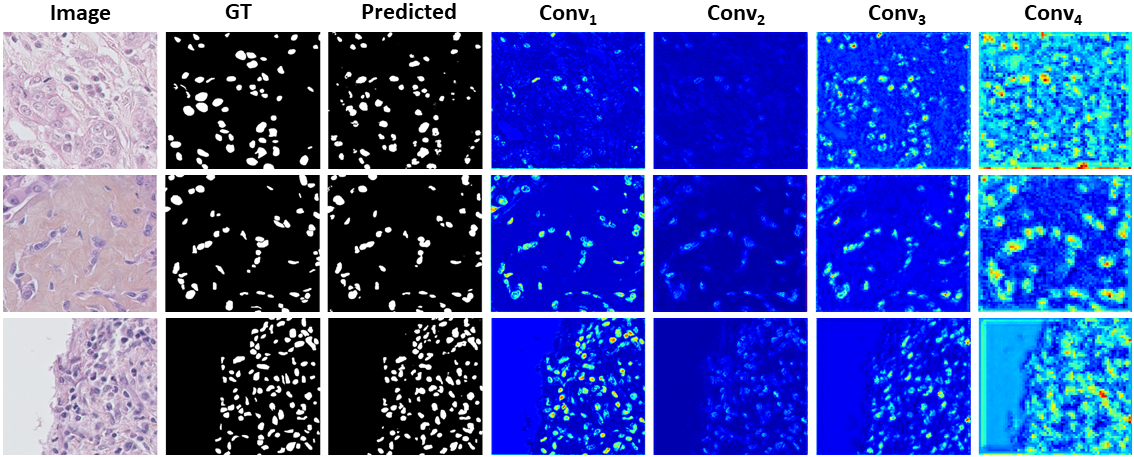}}
     \hfill
      \subfloat[Heatmaps for Encoder and Decoder layers \label{fig:seg_heat_TNBC}]{%
      \includegraphics[width=0.49\textwidth, height=3.5cm]{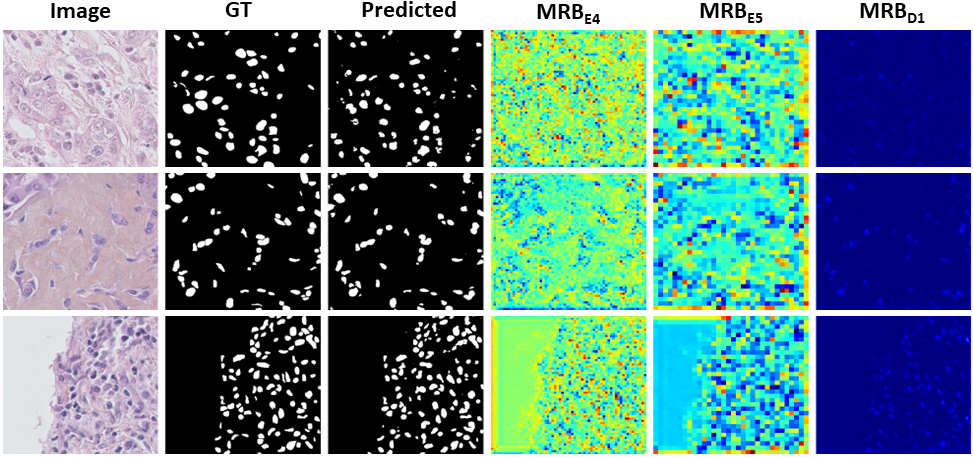}}
      \caption{Segmentation results on the TNBC dataset. GT represents ground truth.}
     \label{fig:TNBC}
\end{figure*}

Table~\ref{ablation1} highlights the substantial impact of controlled information transfer from the encoder side to the decoder side via the skip connections using CDRB. Also, GdAM effectively leverages textual embeddings as prior knowledge to guide attention toward relevant histopathological features, resulting in more accurate segmentation outcomes. Fig. \ref{fig:MonuSeg} and \ref{fig:TNBC} show the segmentation results and the impact of each module of the proposed model on both datasets. 

\begin{table}[!ht]
    \centering
    \caption{Performance of the segmentation models. All values are in \%. Bold values indicate superior performance.}
    \begin{tabular}{cccccc}        
        \textbf{Model} & \textbf{Dice} & \textbf{Recall} & \textbf{Precision}  & \textbf{IoU}\\
        \hline
        (i) & 77.74 & 78.79 & 76.97 & 63.72 \\
        (ii) & 78.82 & 85.60 & 73.17 & 65.10 \\
        (iii) & 79.13 & 82.18 & 77.40 & 66.02 \\
        (iv) & \textbf{80.35} & \textbf{84.11} & \textbf{77.03} & \textbf{67.21} \\
        \hline
    \label{ablation1}
    \end{tabular}
\end{table}

\subsection{SOTA Comparison}
We compare our proposed method with state-of-the-art (SOTA) methods on both datasets, MonuSeg and TNBC, and present the results in Table~\ref{sota_monu} and Table~\ref{sota_tnbc}, respectively. Both the table provides an overview of various evaluation metrics, incorporating established image segmentation models including U-Net~\cite{ronneberger2015u}, Attention-UNet~\cite{oktay2018attention} and DIST~\cite{naylor2018segmentation}. Our proposed method surpasses SOTA models, as demonstrated in Table~\ref{sota_monu} and Table~\ref{sota_tnbc}, regarding both Dice score and IoU. For detailed performance insights, our proposed model achieves a Dice score of 80.35\% (+1.25\%) on MonuSeg and 80.24\% (+2.44\%) on TNBC, indicating substantial similarity between ground truth and predicted segmentation masks. Furthermore, the IoU value of 67.21\% (+1.11\%) on MonuSeg and 66.25\% (+2.05\%) on TNBC highlights the model's robustness in accurately outlining regions of interest.

% \begin{table}[htb]
%     %\centering
%     \caption{Performance comparison of the proposed model with SOTA methods. All values are in \%. Bold values indicate superior performance.}
%     \begin{tabular}{|l|l|l|l|l|} 
%         \hline
%         \multirow{2}{*}{\textbf{Model}} & \multicolumn{2}{c|}{\textbf{Monuseg}} & \multicolumn{2}{c|}{\textbf{TNBC}} \\ \cline{2-5}
%                                         & \textbf{Dice} & \textbf{IoU} & \textbf{Dice} & \textbf{IoU} \\ \hline
%         U-Net\cite{ronneberger2015u} & 74.67 & 60.89 & 68.61 & 52.92\\ \hline
%         Attention U-Net\cite{oktay2018attention} & 78.67 & 66.51 & 71.43 & 54.21\\ \hline
%         DIST\cite{naylor2018segmentation} & 77.31 & 63.77 & 70.51 & 56.34\\ \hline
%         MMPSO-S~\cite{kanadath2023multilevel} & 72.00 & 56.00 & 65.00 & 49.00 \\ \hline
%         Deep-Fuzz~\cite{das2023deep} & 79.10 & 66.10 & 77.80 & 64.20 \\ \hline
%         \textbf{Proposed} & \textbf{80.35} & \textbf{67.21} & \textbf{80.24} & \textbf{66.25}\\ \hline
%     \end{tabular}
%     \label{sota}
% \end{table}

\begin{table}[!ht]
    \centering
    \caption{Performance comparison of the proposed model with SOTA methods on the Monuseg dataset. All values are in \%. Bold values indicate superior performance.}
    \begin{tabular}{cccc}        
        \textbf{Model} & \textbf{Year} & \textbf{Dice} & \textbf{IoU}\\
        \hline
        U-Net~\cite{ronneberger2015u} & 2015 & 74.67 & 60.89 \\ 
        Attention U-Net~\cite{oktay2018attention} & 2018 & 78.67 & 66.51 \\ 
        DIST~\cite{naylor2018segmentation} & 2018 & 77.31 & 63.77 \\ 
        MedT~\cite{valanarasu2021medical} & 2021 & 79.55 & 66.17 \\
        HistoSeg~\cite{wazir2022histoseg} & 2022 & 75.08 & \textbf{71.06} \\
        Deep-Fuzz~\cite{das2023deep} & 2023 & 79.10 & 66.10 \\
        SPPNet~\cite{xu2023sppnet} & 2023 & 79.77 & 66.43 \\
        DCSA-Net~\cite{islam2023densely} & 2023 & 73.20 & 58.00 \\
        MMPSO-S~\cite{kanadath2023multilevel} & 2023 & 72.00 & 56.00 \\
        TSCA-Net~\cite{fu2024tsca} & 2024 & 80.23 & 67.13 \\
        \textbf{Proposed} & - & \textbf{80.35} & 67.21 \\
        \hline
    \label{sota_monu}
    \end{tabular}
\end{table}

\begin{table}[!ht]
    \centering
    \caption{Performance comparison of the proposed model with SOTA methods on TNBC dataset. All values are in \%. Bold values indicate superior performance.}
    \begin{tabular}{cccc}        
        \textbf{Model} & \textbf{Year} & \textbf{Dice} & \textbf{IoU}\\
        \hline
        U-Net~\cite{ronneberger2015u} & 2015 & 68.61 & 52.92\\ 
        Attention U-Net~\cite{oktay2018attention} & 2018 & 71.43 & 54.21\\ 
        DIST~\cite{naylor2018segmentation} & 2018 & 70.51 & 56.34\\
        MCFNet~\cite{Feng_2021_ICCV} & 2021 & 73.37 & 57.94 \\
        Deep-Fuzz~\cite{das2023deep} & 2023 & 77.80 & 64.20 \\ 
        CellTranspose~\cite{keaton2023celltranspose} & 2023 & 77.68 & 59.06 \\
        AlexSegNet~\cite{singha2023alexsegnet} & 2023 & 66.88 & - \\
        MMPSO-S~\cite{kanadath2023multilevel} & 2023 & 65.00 & 49.00 \\
        \textbf{Proposed} & - & \textbf{80.24} & \textbf{66.25} \\
        \hline
    \label{sota_tnbc}
    \end{tabular}
\end{table}

Additionally, we have conducted a cross-dataset experiment to test the effectiveness of our model. Specifically, we have trained our model on the TNBC dataset and tested it on the MonuSeg dataset (TNBC-MonuSeg), as well as trained on the MonuSeg dataset and tested it on the TNBC dataset (MonuSeg-TNBC). Table \ref{cross} shows that our GRU-Net model outperforms popular segmentation models like ResU-Net \cite{diakogiannis2020resunet} and DeepLabv3+ \cite{kuo2022using} in the TNBC-MonuSeg setup while delivering comparable results in the MonuSeg-TNBC setup.

\begin{table*}[hbt]
    \centering
    \caption{Performance comparison of the proposed model with SOTA methods for cross-dataset setups. All values are in \%. Bold values indicate superior performance.}
    \begin{tabular}{|l|l|l|l|l|} 
        \hline
        \multirow{2}{*}{\textbf{Model}} & \multicolumn{2}{c|}{\textbf{TNBC-MonuSeg}} & \multicolumn{2}{c|}{\textbf{MonuSeg-TNBC}} \\ \cline{2-5}
                                        & \textbf{Dice} & \textbf{IoU} & \textbf{Dice} & \textbf{IoU} \\ \hline
        Attention U-Net\cite{oktay2018attention} & 65.21 & 48.60 & 51.41 & 35.65\\ \hline
        ResU-Net\cite{diakogiannis2020resunet} & 64.49 & 47.81 & 67.58 & 51.21\\ \hline
        DeepLabv3+\cite{kuo2022using} & 63.68 & 46.76 & \textbf{73.20} & \textbf{57.76}\\ \hline
        \textbf{Proposed} & \textbf{65.98} & \textbf{49.33} & 71.74 & 54.41\\ \hline
    \end{tabular}
    \label{cross}
\end{table*}

\subsection{Analysis of error cases}
\begin{figure}[!ht]
\centering
\includegraphics[width=\linewidth]{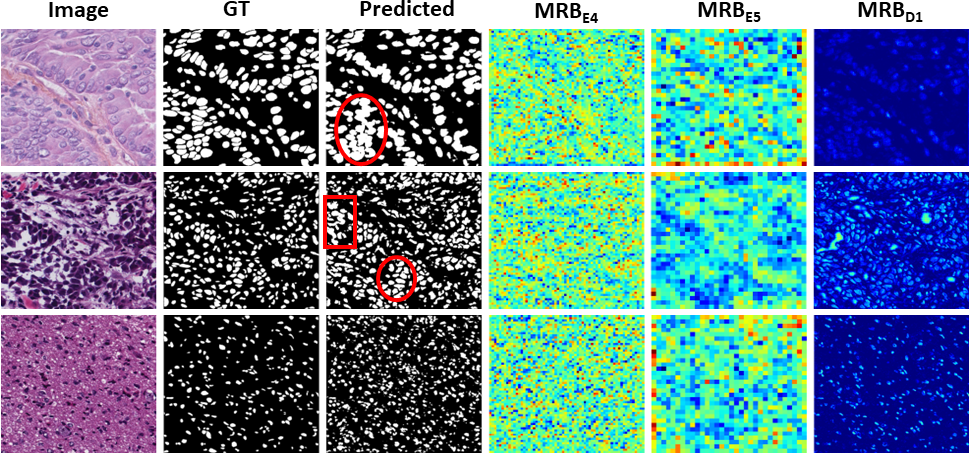}
\caption{Some error cases as produced by the proposed model.}
\label{fig:error}
\end{figure}

Although the proposed model has achieved SOTA segmentation results on two challenging histopathology image datasets, some limitations need to be addressed. In particular, the model faces some challenges in accurately segmenting certain regions marked by red circles and rectangles in the first and second images, as shown in Fig. \ref{fig:error}. To improve the segmentation quality, we could consider refining the boundary constraints of these regions to make them more precise and sharp. Additionally, in the third image, there is an issue of over-segmentation due to the complexity of the image, which contains various noisy points that resemble foreground pixels with similar intensity and color distribution. To tackle this specific issue, augmentation like stain-normalization could be explored in this field. Also, it is necessary to provide textual information for each patch image, even during inference. For this work, such textual information is provided at the whole slide image (WSI) level rather than at the patch level.

\section{Conclusion}
In this paper, we have developed a new model, called GRU-Net, which is used for segmenting nuclei in histopathology images. The model has two main modules: GdAM and CDRB. The CDRB module controls the transfer of information between the encoder and the decoder layers, whereas the GdAM module uses a multi-modal attention mechanism to incorporate relevant text information and highlight specific spatial regions based on prior knowledge. To achieve SOTA results, we have integrated these components into the MultiResU-Net backbone. We have also ensured the robustness of the proposed model using a cross-dataset experimental setup.  

However, we acknowledge that there is always room for improvement in medical applications. Currently, our GdAM module is limited to histopathology image segmentation due to the incorporation of histopathology-relevant information. However, we plan to expand it to other domains and explore text prompts relevant to multiple biomedical segmentation modalities while also focusing on reducing the number of trainable parameters to improve computational efficiency and reduce latency.

\bibliographystyle{splncs04}
\bibliography{Manuscript}
\end{document}